\documentclass[aps,pra,epsf,superscriptaddress,amsmath,amssymb,amsfonts,twocolumn,showpacs,nofootinbib,longbibliography]{revtex4-1}

\usepackage{epsfig}
\usepackage{dcolumn}
\usepackage{bm}
\usepackage{braket}
\usepackage{amsmath}
\usepackage{mathtools}
\usepackage{graphicx,color,xcolor} 
\usepackage{multirow}
\usepackage{lipsum}
\usepackage{etoolbox}
\usepackage[T1]{fontenc}
\usepackage[colorlinks=true,
            linkcolor=red,
            urlcolor=blue,
            citecolor=blue]{hyperref}

\usepackage[normalem]{ulem} 

\begin{document}

\title{Suppressed correlation-spreading in a one-dimensional Bose-Hubbard model with strong interactions}

\author{Jose Carlos Pelayo}
\email{josepelayo@phys.kindai.ac.jp}

\author{Ippei Danshita}
\email{danshita@phys.kindai.ac.jp}
\affiliation{Department of Physics, Kindai University, Higashi-Osaka, Osaka 577-8502, Japan}

\date{\today}

\begin{abstract} 

We investigate signatures of non-ergodic behavior in the real-time evolution of a one-dimensional Bose-Hubbard model, where the initial state is a doubly occupied density-wave state. 
We show that the occupation dynamics at strong interactions is dominated by doublon-holon exchange which leads to a domain wall excitation and propagation. The latter manifests as a negated staggered pattern in the density-density correlations. While the single-particle and the pair correlation functions show highly localized correlations that decay rapidly away from the nearest neighbor. 
We show that the time scale of the domain-wall excitations depends on the inverse of the interaction strength and therefore dictates the slow relaxation dynamics. 
In the presence of a parabolic trap, the occupation dynamics at the edges become frozen and further suppresses the propagation of correlations. This suppression happens even for trap strengths weaker than the tunneling rate. 
We also show that the model can be mapped to an antiferromagnetic transverse-field Ising model in the limit of strong interactions and that the correlation-propagation velocity in the original model is well captured by the group velocity of the spin-wave excitation in the effective spin model.
\end{abstract}

\maketitle

\section{Introduction} 
The ergodicity principle plays a crucial role in connecting the macroscopic theory of thermodynamics and microscopic theory of statistical mechanics. It states that for a system in equilibrium, the long time average of an observable matches that of the ensemble average~\cite{kardar2007statistical,pathria2017statistical}. Isolated many-body quantum systems, like their classical counterparts, can also thermalize despite undergoing a unitary evolution. This occurs when the eigenstate thermalization hypothesis (ETH) is satisfied, the strong version of which states that the long time expectation value of an observable with respect to any single eigenstate equals the ensemble average in a small interval around the eigenenergy \cite{deutsch1991quantum,srednicki1994chaos}. 
There are, however, cases where quantum systems break the strong ETH. 
Such violations  typically arise  due to the presence of an extensive number of conserved quantities.  Prominent examples include integrable systems~\cite{kinoshita2006quantum,rigol2007relaxation,Rigol2008,rigol2009breakdown} and many-body localized (MBL) systems~\cite{ros2015integrals,altman2015universal,Nandkishore2015,Abanin2019,Sierant2025}. In other instances, thermalization is avoided only for a subset of states, such as quantum many-body scar states (QMBS) and states that have a large overlap with QMBS, while the rest of the spectrum remains thermal~\cite{turner2018weak,serbyn2021quantum,moudgalya2022quantum,gotta2023asymptotic,Chandran2023}. 
In general, constraints are present in a non-ergodic system that prevents the exploration of the entire Hilbert space and thus breaks ETH~\cite{moudgalya2022quantum,d2016quantum}. A distinct manifestation of this phenomenon occurs in Hilbert space fragmented (HSF) systems~\cite{moudgalya2022quantum,de2019dynamics,sala2020ergodicity}.  Here, the Hilbert space splits into an exponentially large number of disconnected sectors due to the presence of dynamical constraints, thereby preventing thermalization. 


Cold atoms, owing to their versatility in terms of state preparation, measurement and parameter control~\cite{bloch2012quantum,Schafer2020,Browaeys2020,Daley2022},
have been employed to investigate thermalization and ergodicity breaking across a wide range of many-body realizations including, bosonic, fermionic, and spin systems~\cite{kinoshita2006quantum,Gring2012,trotzky2012probing,Islam2015,schreiber2015observation,bordia2016coupling,choi2016exploring,luschen2017observation,Bernien2017,Lukin2019,Rispoli2019,Scherg2021,Kohlert2023,Wadleigh2023,Su2023,Adler2024,Zhao2025,honda2025observation}.
In particular, Bose-Hubbard systems, which consist of ultracold Bose gases in optical lattices~\cite{Jaksch1998}, have been extensively studied in relation to thermalization and ergodicity~\cite{trotzky2012probing,Islam2015,Lukin2019,Rispoli2019,Wadleigh2023,Su2023,Adler2024,Zhao2025,honda2025observation}.
It has been shown that in the one-dimensional (1D) Bose-Hubbard systems, when the initial state of time evolution is a singly-occupied density-wave state, the density relaxes quickly to equilibrium values ~\cite{trotzky2012probing}. However, in the presence of a tight trap, the particles become localized along the trap edges and the system exhibits stark MBL~\cite{yao2020many,yao2021nonergodic}.
On the other hand, it has been shown that dynamics starting from a doubly occupied density-wave state also becomes non-ergodic due to the combined effects of a weak trapping potential and strong interatomic interaction~\cite{honda2025observation,kunimi2021nonergodic}. In these works signatures in the eigenenergies, i.e., level statistics as well as dynamical signatures in some physical quantities including the entanglement entropy and the particle-number imbalance have been used to detect the effects of non-ergodicity.

In this paper, focusing on the dynamical spreading of quantum correlations, we further explore signatures of non-ergodicity of the 1D Bose-Hubbard system in the quench dynamics starting from a doubly occupied density-wave state. 
Correlation-spreading dynamics has attracted much interest mainly for understanding mechanisms of quantum information propagation and thermalization~\cite{Gong2022,Cheneau2022}. Previous experiments have addressed such dynamics of the Bose-Hubbard systems taking a Mott insulator with unit filling as the initial state and observed the light-cone-like propagation in the density-density~\cite{cheneau2012light} and single-particle~\cite{takasu2020energy} correlation functions. The dynamics in the similar settings have been theoretically analyzed in many previous studies~\cite{Lauchli2008,Barmettler2012,Natu2013,Carleo2014,Fitzpatrick2018,Nagao2019,Despres2019,Nagao2021,Mokhtari-Jazi2021,Kaneko2022,Nagao2025}. Moreover, it has been shown in Ref.~\cite{Rispoli2019} that when a quasi-periodic potential is added to the 1D Bose-Hubbard system, the correlation spreading is significantly suppressed in the parameter region of the many-body localization. These experimental results motivate us to address a question of whether a similar suppression of the correlation spreading occurs also in the case of the dynamics starting from a doubly occupied density-wave state.

We present quasi-exact numerical simulations of the real-time dynamics of the 1D Bose-Hubbard model with use of matrix-product states (MPS)~\cite{Schollwock2011}. We first identify the relevant excitations by examining the site-occupation dynamics and contrast their behavior in the presence of a trapping potential. 
We then show that the correlation spreading is suppressed for finite trap strengths. In the absence of a trap, however, depending on the type of the correlator, its evolution may show either localization or relaxation at intermediate time scales.
To gain deeper insight into the dynamics, we also derive an effective Hamiltonian that maps the system to a transverse-field Ising (TFI) model which provides an intuitive way to understand the underlying dynamics. 

This paper is organized as follows: In Sec.~\ref{sec:Model} we specify our model as well as the methods. In Sec.~\ref{sub_sec: occ_dynamics} we examine the site-occupation dynamics in an open boundary system for both vanishing and finite trap. This is followed by investigating different types of correlators in Sec.~\ref{sub_sec:correlations} and the introduction of an effective model in Sec.~\ref{sub_sec:eff_models}. We then conclude our work in Sec.~\ref{sec:Conclusions}. Finally, we include the appendix~\ref{app:Map_to_TFI} which detail the derivation of the effective model.

\section{Model and Methods}\label{sec:Model} 

We consider a 1D Bose-Hubbard chain with open boundary condition, under the influence of a parabolic trapping potential. The Hamiltonian is defined as

\begin{equation}
    \hat{H} = -J\sum_{\langle i,j\rangle}\hat{a}_i^\dagger\hat{a}_j + \frac{U}{2}\sum_{i=1}^M \hat{n}_i(\hat{n}_i - 1) + \Omega\sum_{i=1}^M V_i\hat{n}_i, \label{eq:Ham_main}
\end{equation}
\noindent
where $\hat{a}_i^\dagger$ $(\hat{a}_i)$ is the bosonic creation (annihilation) operator at site $i$, while $\langle i,j\rangle$ indicates a summation over the nearest-neighbor sites with a hopping amplitude of $J > 0$.  The interactions are assumed to be on-site with strength $U > 0$ and $\hat{n}_i=\hat{a}_i^\dagger \hat{a}_i$ being the occupation number operator at site $i$. The trapping potential of strength $\Omega \ge 0$ is defined as $V_i = (i - (M+1)/2)^2$  which preserves reflection symmetry. In the following, all parameters of the system will be scaled by the hopping amplitude $J$ and we set the lattice spacing to unity.

As discussed in the previous section, we are interested in the dynamics starting from a doubly occupied density-wave state which can be represented in the occupation number basis as 
$\ket{\psi_0} = \ket{\nu_1,\nu_2,\nu_3...}= \ket{2\,0\,2\,0...}$ with $\nu_i$ being the occupation number of site $i$. 
This initial state can be prepared, for example, in cold atom experiments by tuning the relative lattice depths and site offsets of a superlattice potential, thereby redistributing particles~\cite{honda2025observation,trotzky2012probing,lohse2016thouless}. Since the initial state, $\ket{\psi_0}$, is not an eigenstate of the Hamiltonian, the evolution of the system highly depends on the interaction strength and the trapping potential. For weak interactions, $U/J \ll 1$, the system quickly thermalizes into a state with one atom per site on average~\cite{carleo2012localization,russomanno2020nonergodic}. On the other hand, for strong interactions, $U/J \gg 1$, the thermalization slows down preserving the density wave for a longer time, while the effect of the parabolic trapping potential further slows down the thermalization~\cite{kunimi2021nonergodic,honda2025observation}. The latter response of the system is a manifestation of ETH-breaking due to Hilbert space fragmentation which can be quantified through the imbalance measuring the difference in the particle number in the odd and even sites. This quantity yields unity for the density-wave state and vanishes for a thermal state. In this work, we are interested in other signatures of ETH-breaking such as in the correlation propagation, and thus we will be focusing on parameter regimes that yield such a response.    

All calculations in this work, unless otherwise stated, are done using a tensor network formalism in an open boundary condition. The time evolution of the state in the MPS form, is computed using a time-evolving block decimation (TEBD) algorithm~\cite{Schollwock2011,Vidal2004} with a maximum bond dimension of $\chi_{\text{max}} = 1600$. Moreover, we restrict the maximum occupation per site up to $n_{\text{max}} = 4$, which is valid for large $U/J$. The lattice size is set to $M=16$ ($M=12$) with final evolution times of $t= 10^2 \hbar/J$ ($t= 10^3 \hbar/J$). In all calculations where the trapping potential is finite, we used a weak trap with strength $\Omega/J = 0.02$ in order to minimize its direct influence, thus ensuring the observed dynamics arise mainly from interparticle interactions.

\section{Results}\label{sec:Results} 
\subsection{Site occupation dynamics \label{sub_sec: occ_dynamics}}
\begin{figure}[tb]
\centering  
\includegraphics[width=\linewidth]
{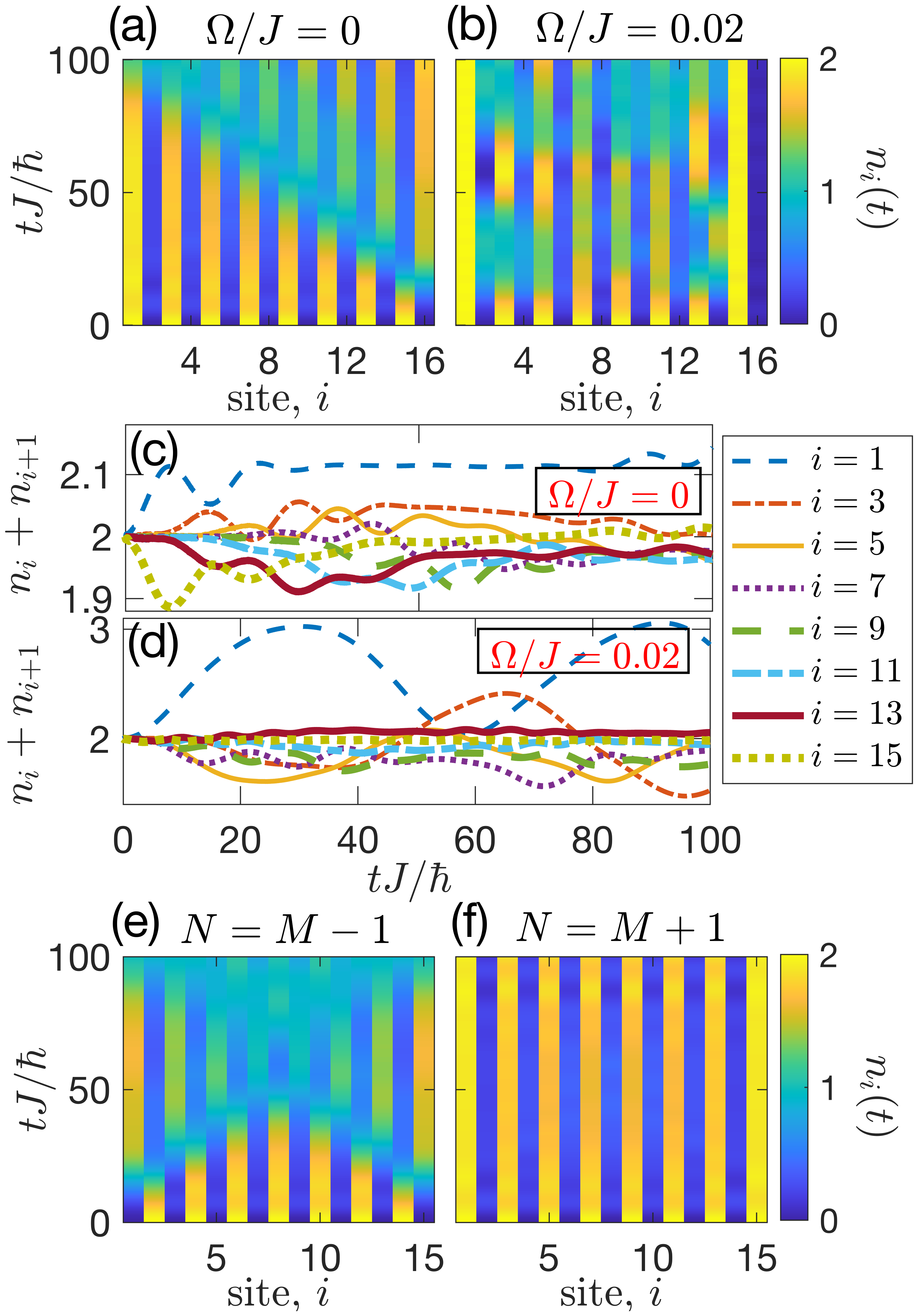}
\caption{Time-evolved site occupations at strong interaction $U/J = 40$. For $M=16$, results are shown for $\Omega/J=0$ (a) and $\Omega/J=0.02$ (b), with the corresponding neighboring two-site occupations in (c) and (d). Panels (e) and (f) show the site occupation for  $M=15$ with particle number $N=M-1$ and $N=M+1$, respectively.}
\label{fig:site_occupation}
\end{figure}

When $U\gg J$ the initial state $\ket{\psi_0}$  is a high-energy state of the Hamiltonian in Eq.~\eqref{eq:Ham_main} due to its doubly-occupied density wave configuration. A single-particle hopping process from $\ket{\psi_0}$ dissociates a doublon into two nearest-neighboring singlons, yielding an energy decrease proportional to $\sim U$ and is therefore suppressed. In this sense the doublon is often called a repulsively bound pair~\cite{Winkler2006}. Instead of the first-order processes with respect to the single-particle hopping, the second order processes such as pair tunneling~\cite{carleo2012localization} becomes dominant as we shall see in the following.

We begin by examining the time evolution of the site occupation, $n_i = \braket{\hat{n}_i}$, in the case of a vanishing trapping potential, $\Omega/J = 0$, and a weak trapping potential, $\Omega/J = 0.02$, at moderately strong interactions ($U/J = 40$) shown in Fig.~\ref{fig:site_occupation}. For $\Omega/J = 0$, the initial site-occupation of the density wave is maintained until $tJ/\hbar \sim U/2J$. Thereafter, the rightmost edge with an empty site is filled by its neighbor through a pair hopping process leaving the site next to the rightmost edge empty. This pair hopping process then cascades to the rest of the system (see Fig.~\ref{fig:site_occupation} (a)) flipping a site pair from $\ket{2 \,0}\rightarrow\ket{0\,2}$ until it reaches the opposite edge where the pair hopping inverts the system back to the original configuration~\cite{weckesser2025realization}. The pair hopping or doublon-holon exchange is also apparent from the two-site occupation of neighboring sites, $n_i+n_{i+1}$, which shows small fluctuations for large $U$ (see Fig.~\ref{fig:site_occupation} (c)). 

The pair hopping dynamics can be interpreted as a domain wall being excited at the empty edge which then propagates to the opposite edge. Here, the domain wall separates the $\ket{2\,0}$ pairs from the $\ket{0\,2}$ pairs. 
In the case where both edges are empty, as in the case of $N=M-1$ with $N$ being the particle number, the domain wall propagates from both sides (see Fig.~\ref{fig:site_occupation} (e)). On the other hand, in the case of $N=M+1$, the domain wall is not excited since there is no empty edge (see Fig.~\ref{fig:site_occupation} (f)). 
This boundary dependence of the dynamics comes from the difference in energy cost of the pair hopping process resulting in two adjacent empty sites or two adjacent doubly occupied sites. The former is preferred with |$\Delta E| \sim 2J^2/U$ when there is an initial empty edge but is not present when there is no empty edge, while the latter's energy cost is |$\Delta E| \sim 16J^2/U$ (see effective Hamiltonian (Eq.~\eqref{eq:eff_BH}) in Sec.~\ref{sub_sec:eff_models} for a clearer picture). 
Once the domain wall is excited, the system begins to thermalize as it starts to lose its original density-wave configuration. At the same time, the imbalance parameter also begins its decay towards zero imbalance signifying a relaxation process (not shown here). 

In the presence of a trap, even for weak ones ($\Omega/J = 0.02$), such as in Fig~\ref{fig:site_occupation} (b), the domain wall propagation is replaced by a more complicated particle dynamics as the edges become energetically separated from the rest of the system and thus are localized as compared to the inner sites. 
In fact, for $U/J \gg 1$, the pair hopping at the left edge costs the largest energy and therefore fixes $n_1(t)$. 
As a result, the two-site occupation in Fig.~\ref{fig:site_occupation}(d) shows large fluctuations for the pairs close to the left edge. On the right edge, however, fluctuations are smaller as the occupation of the rightmost pair is fixed. This asymmetric behavior comes from the combined effect of a symmetric trap but with asymmetric initial state. 

The effect of increasing interaction strengths on the site occupation dynamics, in the case of $\Omega/J = 0$, is to modify the time scales and the excitation propagation speed relative to an effective hopping strength $\tilde{J}=2J^2/U$ (see section~\ref{sub_sec:eff_models}). In addition, for finite trapping strengths, as $U/J$ increases, more and more outer sites away from the center have a fixed occupation and significant changes in $n_i(t)$ only occur close to the center of the chain (not shown here). Note that this behavior is not due to the presence of a strong trap, but rather to the fact that the dynamics is now governed by an effective hopping strength $\tilde{J}$ that becomes weaker for larger $U$.

\begin{figure*}[tb]
\centering  
\includegraphics[width=0.9\linewidth]
{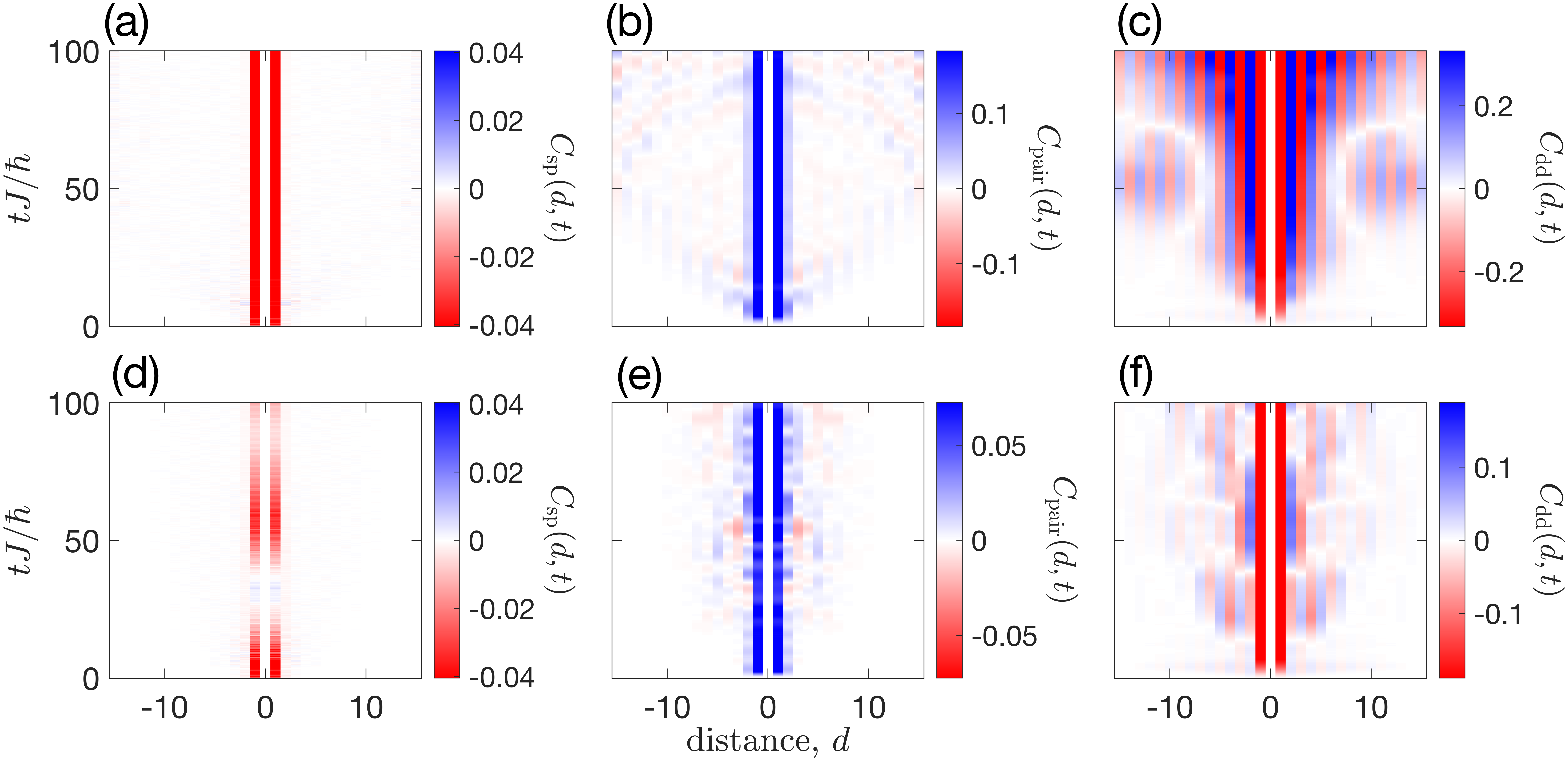}
\caption{Evolution of different correlation functions (see colorbar label) for $\Omega/J = 0$ (upper row) and $\Omega/J = 0.02$ (lower row). In all cases, the interaction strength is $U/J = 40$ and $M=16$ while the correlations at $d=0$ are not shown. }
\label{fig:corr_fucntions}
\end{figure*} 

\subsection{Correlations \label{sub_sec:correlations}}

In both cases shown above, it is apparent that thermalization can be suppressed  at relatively long times for $U/J \gg 1$. It is then interesting to see whether such a signature is also present in the time-evolved correlations. To investigate this, we compute several equal-time correlation functions defined as

\begin{subequations}
\begin{align}
    C_{\text{sp}}(d,t) = \frac{1}{N_c}\sum_{i,j=i+d} \Re\left[\braket{\hat{a}_i^\dagger(t)\hat{a}_j(t)}\right], \label{eq:corr_sp}\\
    C_{\text{pair}}(d,t) = \frac{1}{N_c}\sum_{i,j=i+d} \Re\left[\braket{\hat{a}_i^\dagger(t)\hat{a}_i^\dagger(t)\hat{a}_j(t)\hat{a}_j(t)}\right], \label {eq:corr_pair}\\
    C_{\text{dd}}(d,t) = \frac{1}{N_c}\sum_{i,j=i+d} \braket{\hat{n}_i(t)\hat{n}_j(t)}-\braket{\hat{n}_i(t)}\braket{\hat{n}_j(t)}, \label{eq:corr_density}
\end{align}
\end{subequations}
where $C_{\text{sp}}(d,t)$, $C_{\text{pair}}(d,t)$, and $C_{\text{dd}}(d,t)$ are the single-particle correlation function, pair correlation function, and the connected density-density correlation function, respectively. Here, $N_c$ is a normalization constant that counts the number of $i,j$-pairs for a given distance $d=j-i$ and the symbol $\Re$ denotes the real-part of the correlators.  
    
The evolution of the single-particle correlation function in 
 Fig.~\ref{fig:corr_fucntions}(a) shows a highly-localized behavior with the correlations being significant only in the nearest-neighbor. This is in contrast to the ballistic spreading of correlations observed in the case of a quench from an initial Mott state to that close to the Mott-superfluid transition~\cite{cheneau2012light,takasu2020energy}. The correlation strength of $C_{\text{sp}}$ is about an order of magnitude smaller compared to the pair correlation function $C_{\text{pair}}$ (see Fig~\ref{fig:corr_fucntions}(b)). This is to be expected since the occupation dynamics of the system is dominated by the pair hopping in the large $U/J$ regime. Although the $C_{\text{pair}}$ shows a bit more correlation spreading, the nearest neighbor correlation still dominates and thus signals a localization of $C_{\text{pair}}$. The trap effects on both $C_{\text{sp}}$ and $C_{\text{pair}}$ reduce the overall strength of the correlations and also restrict the weak correlation spreading to a smaller distance in the latter case (see Figs.~\ref{fig:corr_fucntions}(d) and (e)). In contrast to the clear signatures of localization in the two correlators discussed above, the density-density correlation function maintains a non-negligible correlations even at long distances (see Fig~\ref{fig:corr_fucntions}(c)). This is attributed to the persisting density-wave pattern of the time-evolved state which can be seen from the alternating positive to negative correlations for even to odd distances. As the state evolves from an initial product state, the state occupations of other density-wave-like states grow along with the correlation spreading $C_{\text{dd}}$. 
 Notice that at $tJ/\hbar \sim U/2J$ and at relatively large distances the $C_{\text{dd}}$ displays a negated staggered pattern relative to the dominant pattern. This is a signature of the domain wall excitation and propagation that we see in Fig.~\ref{fig:site_occupation}(a). 
 In spite of this perceivable spreading, the nearest-neighbor correlation dominates and thus signals a weak localization. By introducing the trap, the overall strength of the correlations is reduced while keeping the staggered correlations (see Fig~\ref{fig:corr_fucntions}(f)). In addition, the correlation spreading is now restricted to smaller distances which means that the suppression is further enhanced. 

\begin{figure}[tb]
\centering  
\includegraphics[width=.9\linewidth]
{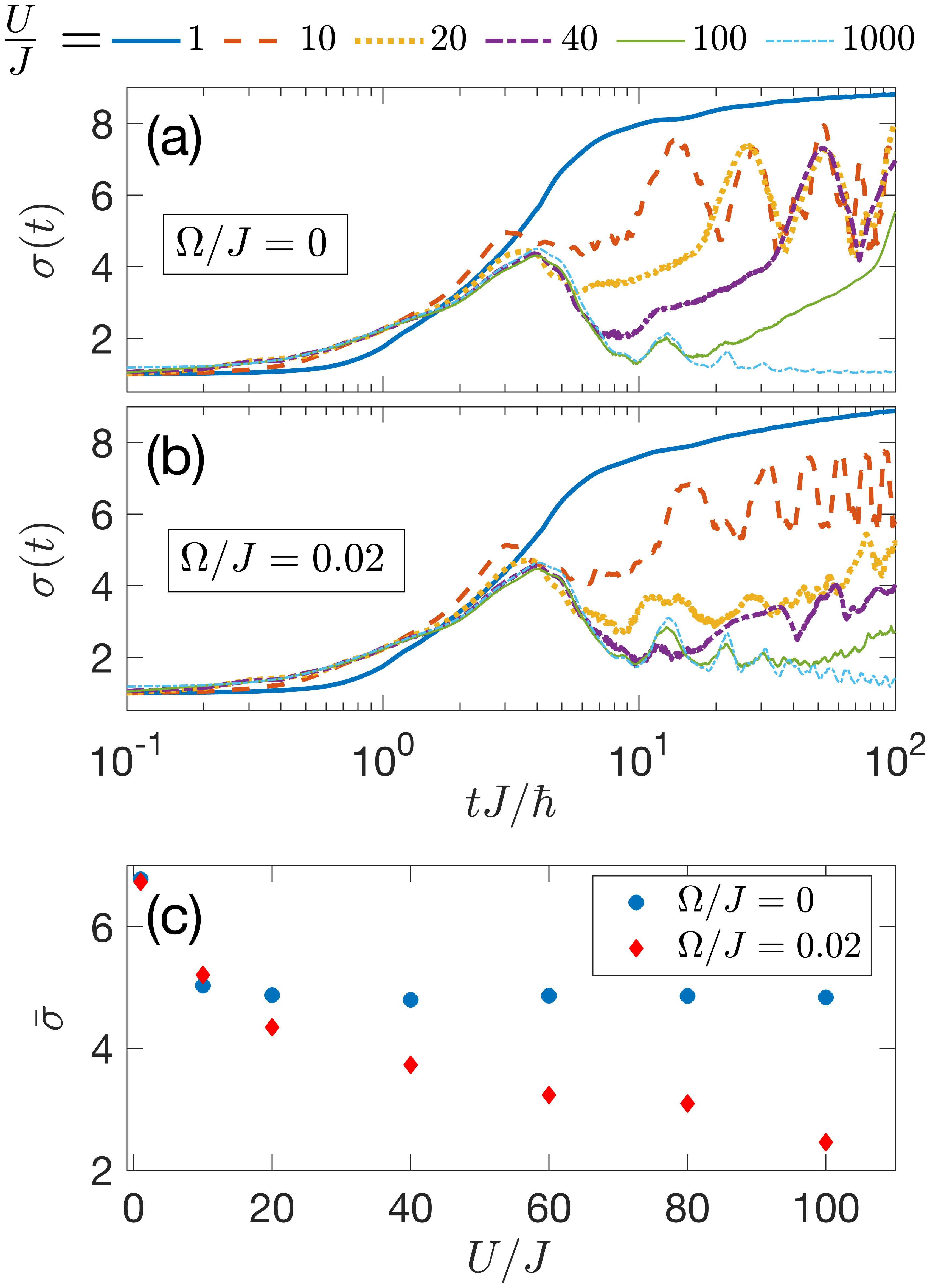}
\caption{Correlation width as a function of increasing evolution time for (a) $\Omega/J = 0$ and (b) $\Omega/J = 0.02$, where $M=16$. (c) Long-time average of correlation width as a function of increasing interactions for different trap strengths (see figure legend), where $M=12$.}
\label{fig:corr_width}
\end{figure}

 To better quantify the spreading of the correlations in $C_{\rm dd}$, we compute the standard deviation of the width~\cite{de2019dynamics} defined as,
 \begin{equation}
     \sigma(t) = \left[\frac{1}{N_\sigma}(t)\sum_d d^2 C_{\rm dd}(t)\right]^\frac{1}{2},
 \end{equation}
\noindent
where $N_\sigma(t) = \sum_d C_{\rm dd}(d,t)$ is the normalization constant. Note that the $\sum_d d\,C_{\rm dd}(d,t)$ vanishes. $\sigma(t)$ can be interpreted as the width at which the correlations have spread from any point in the lattice (since the quench is global)~\cite{magoni2024coherent}. For weak interactions $U/J = 1$, regardless of the presence of a trap, a saturation tendency of $\sigma(t)$ can be seen in Figs.~\ref{fig:corr_width}(a) and (b) signifying that the state has  thermalized. On the other hand, for $U/J \gg 1$ and $\Omega/J = 0$, the first oscillation peak of $\sigma(t)$ is shifted to longer times for increasing $U/J$, implying that thermalization can be suppressed as $U/J$ is increased.  This behavior comes from the effective tunneling coefficient, $\tilde{J}=2J^2/U$, being proportional to $U^{-1}$. For $\Omega/J = 0.02$, the oscillatory behavior is still present but now there is a gradual reduction of $\sigma(t)$ as $U$ is increased. This behavior captures the suppression of correlation spreading to smaller distances as also seen in Fig.~\ref{fig:corr_fucntions}(f).

To see whether the perceived suppression of the correlation persists at much longer time, we take the long time average  $\bar{\sigma}(t)=(t_f-t_i)^{-1}\int^{t_f}_{t_i} dt \, \sigma(t)$ for a system size of $M=12$ and for $t_i = 10^1\hbar/J$ and $t_i = 10^3\hbar/J$. We see from Fig.~\ref{fig:corr_width}(c) that $\bar{\sigma}$ tends to saturate as a function of increasing $U$ for vanishing $\Omega/J$. Whereas for finite $\Omega/J$, $\bar{\sigma}(t)$ continues to decay for larger $U$. The behavior of $\bar{\sigma}(t)$ in the former implies that the system is in the relaxation regime while in the latter the system remains non-ergodic.

\subsection{Effective Models} \label{sub_sec:eff_models}
To better understand the dynamics of the system, we introduce here effective models that map the bosonic system into spin systems. As seen in the previous sections, the dynamics is dominated by a pair hopping process in the limit of $U\gg J$. In this limit, the trapped Bose-Hubbard model in Eq.~\eqref{eq:Ham_main} can be mapped to a system with a reduced Hilbert space where the occupation number is restricted to only $\nu_i = \{0,2\}$. By applying the Schrieffer-Wolff transformation one obtains the effective model~\cite{kunimi2021nonergodic},

\begin{align}
\hat{H}_{\text{EBH}} &= \sum_i \Big[
J^+_i\ket{0_i2_{i+1}}\bra{0_i2_{i+1}}+J^-_i\ket{2_i0_{i+1}}\bra{2_i0_{i+1}} 
 \nonumber \\
&  +\bar{J}_{i}\left(\ket{0_i2_{i+1}}\bra{2_i0_{i+1}}+h.c. 
-6\ket{2_i2_{i+1}}\bra{2_i2_{i+1}}\right) \Big] \nonumber   \\
&+ \Omega\sum_{i=1}^M V_i\hat{n}_i, \label{eq:eff_BH}
\end{align}
\noindent
where $\bar{J}_{i}, J^+_i$ and $J^-_i$ are site-dependent tunneling coefficients arising from the inhomogeneity caused by the trap. These site-dependent coefficients are all equivalent to $2J^2/U$ in their first-order Taylor series expansion while the sub-leading terms are in the order of $\mathcal{O}(\Omega(V_{i+1}-V_i)/U)$. For our chosen parameter values, the sub-leading terms are $\Omega(V_{i+1}-V_i)/U \lesssim 10^{-2}$ and therefore can be omitted for dynamis at relatively short times. It has been shown that Eq.~\eqref{eq:eff_BH} is equivalent to the spin-$1/2$ XXZ Hamiltonian with a trap~\cite{carleo2012localization,kunimi2021nonergodic}. 

Here, we employ a further mapping from a two-site occupation state to a single Ising spin state as $\ket{2\,0} \rightarrow \ket{\uparrow}$ and $\ket{0\,2} \rightarrow \ket{\downarrow}$. This leads to an effective transverse-field Ising model with a trap and an additional boundary term (see Appendix~\ref{app:Map_to_TFI} for the exact mapping of the Hamiltonian and observables),
\begin{align}
    \hat{H}_{\text{TFI}} = &\tilde{J}\sum_i^{M'}\hat{\sigma}^x_i + 2\tilde{J}\sum_i^{M'-1}\hat{\sigma}^z_i\hat{\sigma}^z_{i+1} +\frac{3\tilde{J}}{2}\left(\hat{\sigma}^z_1 - \hat{\sigma}^z_{M'}\right)  \nonumber \\
    &+ \Omega\sum_{i=1}^{M'} (V_{2i} - V_{2i-1})\hat{\sigma}_i^z, \label{eq:TFI}
\end{align}
\noindent
where $\hat{\sigma}^{x,z}_i$ are Ising spin operators and $\tilde{J} = 2J^2/U$ is the effective spin-flip coefficient. Here, the lattice size, $M'=M/2$, is half the original model as we map a two-site occupation state into a single spin state. This mapping is valid for $U\gg J$ and for vanishing trap strengths where only small fluctuations in the two-site occupation is observed and therefore the mapping can be safely applied (see Fig.~\ref{fig:site_occupation}(c)). On the other hand, for finite $\Omega/J$, the mapping breaks down as the site-occupation at the edge is frozen, yielding a two-site occupation state that cannot be mapped to the Ising states. 
Thus, in the following, we focus only on the case of $\Omega = 0$.

\begin{figure}[tb]
\centering  
\includegraphics[width=\linewidth]
{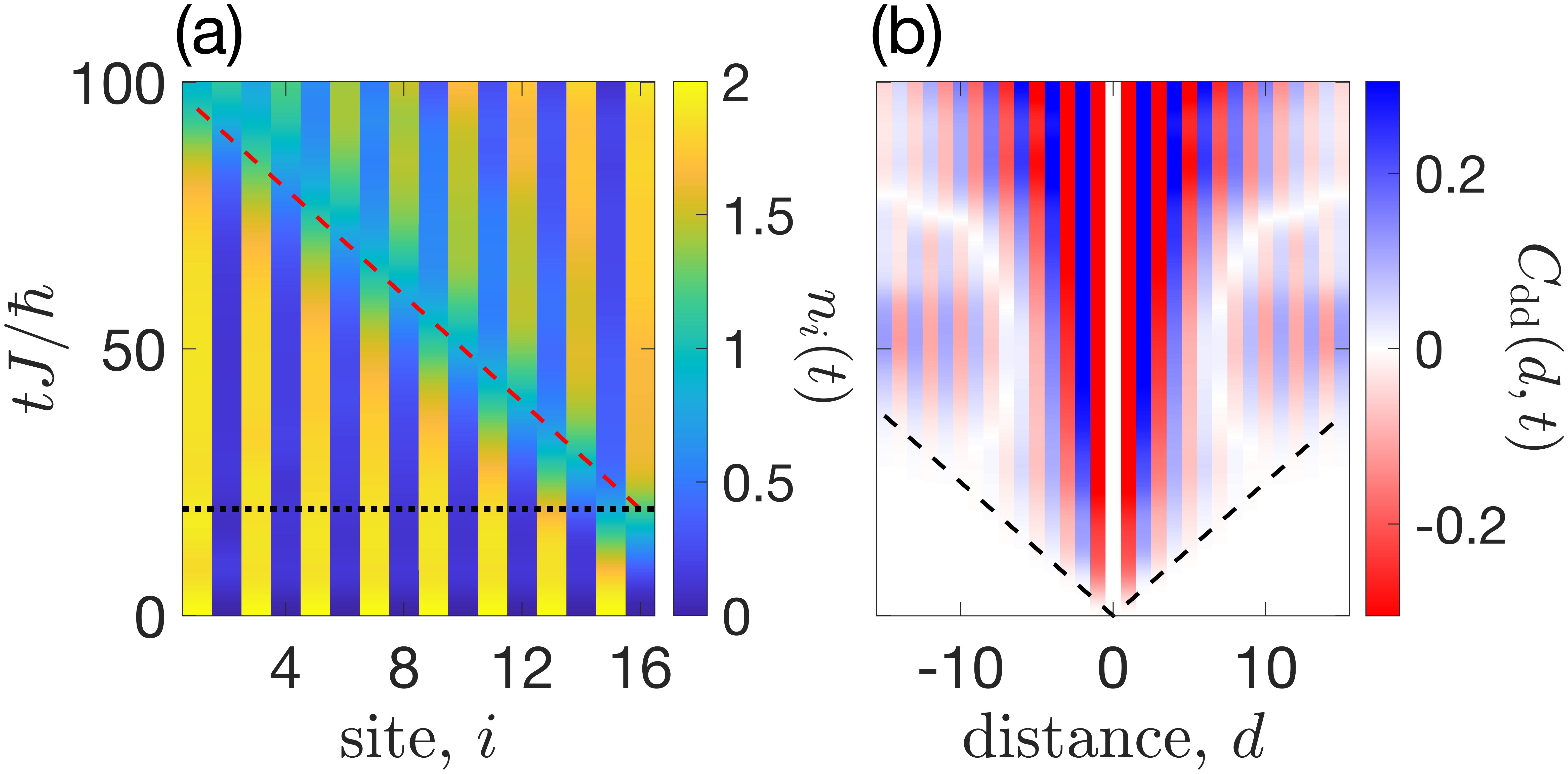}
\caption{(a) Site occupation dynamics and (b) density-density correlation function of the transverse field Ising model obtained by exact time evolution. The dotted line in (a) denotes the time at which the excitation occurs while the dashed lines in (a) and in (b) equals $v_g^{-1}$ and $(2v_g)^{-1}$, respectively. See main text for details.}
\label{fig:TFI_occ_corr}
\end{figure} 

The initial density-wave state can also be represented in terms of the Ising spins as $\ket{\psi_0}=\ket{\uparrow\uparrow\uparrow...}$. Although $\ket{\psi_0}$ is the ferromagnetic ground state, the effective TFI model in Eq.~\eqref{eq:TFI} exhibits an anti-ferromagnetic coupling, therefore, in this representation the initial state is also a high-energy state which evolves non-trivially.
Through the mapping of the Ising spin operators to the occupation number operator $\hat{n}$ (see Appendix~\ref{app:Map_to_TFI}), we compute the site-occupation dynamics using the TFI Hamiltonian and find a good agreement in $n_i(t)$ between the evolution of $\ket{\psi_0}$ under the Bose-Hubbard model and the TFI model (see Fig.~\ref{fig:site_occupation}(a) and Fig.~\ref{fig:TFI_occ_corr}(a), respectively). The previously observed domain wall excitation and propagation can then be interpreted as a spin-flip excitation, $\ket{\uparrow}\rightarrow\ket{\downarrow}$, and propagation in the spin picture. In the latter case, a lower energy cost is needed to excite a spin flip at the right edge due to the imposed boundary term. The coefficient of the spin flip term, $\hat{\sigma}_x$, in Eq.~\eqref{eq:TFI} is then related to the time at which the spin flip excitation occurs being $\sim 1/\tilde{J}$. This is also captured in the occupation dynamics (see dotted line in Fig.~\ref{fig:TFI_occ_corr}(a)). We also find a good agreement in the density-density correlation, $C_{\rm dd}(t)$, obtained from the Bose-Hubbard and the TFI model (see Fig.~\ref{fig:corr_fucntions}(c) and Fig.~\ref{fig:TFI_occ_corr}(b), respectively). The discrepancies in the observables obtained from both models arise from the restriction of the site occupation in the TFI model being only ${n}_i = \{0,2\}$, unlike the Bose-Hubbard model, which allows a broader range of occupations.

One advantage of mapping to a TFI model over the XXZ model is that the excitation spectrum of the former is known and has a closed form, being $\epsilon_k = 4\tilde{J}\left(1 + 1/4 - \cos(2k)\right)^{1/2}$~\cite{Sachdev_2011}. Here, the spectrum refers to the case of $\Omega = 0$ and by omitting the boundary term. The latter condition is justified since we are only interested in the bulk property such as the group velocity that is determined from the spectrum as $v_g = \max \left|\partial\epsilon_k/\partial (\hbar k)\right| = 4\tilde{J}/\hbar$. The dashed line in Fig~\ref{fig:TFI_occ_corr}(a) traces the inverse speed $v_g^{-1}$ and shows good agreement with the observed spin-flip propagation. At the same time, the correlation propagation speed under the TFI model (see dashed line in Fig~\ref{fig:TFI_occ_corr}(b)) also matches well with $2v_g$ as in Ref.~\cite{kaneko2023dynamics}. 

\section{Conclusions}\label{sec:Conclusions} 

We have numerically calculated the real-time dynamics of the 1D Bose-Hubbard model starting from a doubly occupied density-wave state using the matrix-product-state method in order to analyze the slow relaxation that occurs in the strongly interacting regime. We have shown that in the absence of a trapping potential once the domain-wall excitation is formed, relaxation then follows. However, depending on the type of correlator considered, localized correlations may still be observed as in the case of the single-particle and pair correlation functions. In contrast, the density-density correlations support the relaxation findings especially at evolution times much longer than the $\hbar\tilde{J}^{-1}$.
On the other hand, in the presence of a weak trapping potential, slow relaxation becomes more pronounced, leading to suppression of correlation spreading even at long evolution times, regardless of the type of correlators.

We have also demonstrated that mapping to the effective TFI model provides a more intuitive interpretation of the results where the domain wall excitation correspond to spin-flip excitation. The spin model was able to capture the qualitative features of  the site-occupation dynamics and the density-density correlations of the original model. Moreover, the reduction of  the lattice size into half, together with the smaller local dimension in the spin model, enables the exploration of a larger lattice system.  


\section*{Acknowledgments}
We would also like to acknowledge useful discussions with Masaya Kunimi and Yosuke Takasu. The calculations were performed utilizing the TeNPy library \cite{SciPostPhysCodeb}. We acknowledge the support from Quantum Leap Flagship Program from MEXT [Grant No.~JPMXS0118069021], 
FOREST from Japan Science and Technology Agency (JST) [Grant No.~JPMJFR202T], 
and ASPIRE from JST [Grant No.~JPMJAP24C2].  

\appendix

\section{Mapping to Transverse Field Ising model \label{app:Map_to_TFI}}

The mapping from the effective Bose-Hubbard model in Eq.~\eqref{eq:eff_BH} to the TFI model in Eq.~\eqref{eq:TFI} relies on the two-site correspondence $\ket{2\,0} \rightarrow \ket{\uparrow}$ and $\ket{0\,2} \rightarrow \ket{\downarrow}$, together with the leading-order approximation $\bar{J}_i = J^+_i = J^-_i \approx \tilde{J}=2J^2/U$. This is valid for strong interactions and as long as the two-site occupation does not fluctuate too much as in the case of $\Omega/J=0$. We then use the following relations

\begin{align}
    \ket{0_i2_{i+1}}\bra{2_i0_{i+1}}+ h.c. &\rightarrow \hat{\sigma}_j^x \label{eq:TFI_sx} \\
    \ket{2_i0_{i+1}}\bra{2_i0_{i+1}}  &\rightarrow \frac{1}{4}(1+\hat{\sigma}^z_j)(1+\hat{\sigma}^z_{j+1}) \label{eq:TFI_20_02_term}\\
    \ket{0_i2_{i+1}}\bra{0_i2_{i+1}} &\rightarrow \frac{1}{4}(1-\hat{\sigma}^z_j)(1-\hat{\sigma}^z_{j+1})\\
    \ket{2_i2_{i+1}}\bra{2_i2_{i+1}} &\rightarrow \frac{1}{4}(1-\hat{\sigma}^z_j)(1+\hat{\sigma}^z_{j+1}) \label{eq:TFI_22_term},
\end{align}
\noindent
where the index changes from $i =\{1,2,\dots,M\}$ to $j=\{1,2,\dots,M/2\}$ and $\hat{\sigma}^{x,z}_j$ are the Ising spin matrices.  
In addition to Eqs.~\eqref{eq:TFI_sx}-\eqref{eq:TFI_22_term}, an additional term $\tilde{J}\sum_j (\hat{\sigma}^z _j)^2$ must be included for completeness. This term, however, contributes only to an energy shift and is therefore omitted in the final expression of Eq.~\eqref{eq:TFI} along with the constant terms in Eqs.~\eqref{eq:TFI_20_02_term}-\eqref{eq:TFI_22_term}.
The two-spin interaction, such as in  Eq.~\eqref{eq:TFI_22_term}, can be understood more clearly by considering a state like $\ket{ 0\,2\,2\,0}$ or $\ket{ \downarrow\uparrow}$ in the spin picture. In the original model, the interaction between particles in the second and third site is described by Eq.~\eqref{eq:TFI_22_term} in the spin model.

To retrieve the information of the site-occupation operator and the density-density correlation function, we used the following mapping
\begin{align}
    \braket{\hat{n}_{2j}} = 1 - \braket{\hat{\sigma}_j^z} \\
    \braket{\hat{n}_{2j-1}} = 1 + \braket{\hat{\sigma}_j^z} \\
    \braket{\hat{n}_{2i}\,\hat{n}_{2j}}_c = \braket{\hat{n}_{2i-1}\,\hat{n}_{2j-1}}_c = \braket{\hat{\sigma}_i^z\hat{\sigma}_j^z}_c \\
    \braket{\hat{n}_{2i}\,\hat{n}_{2j-1}}_c = \braket{\hat{n}_{2i-1}\,\hat{n}_{2j}}_c = -\braket{\hat{\sigma}_i^z\hat{\sigma}_j^z}_c,
\end{align}
where $i\neq j$ and the subscript in $\braket{..}_c$ denotes a connected correlation, i.e., $\braket{A\,B}_c = \braket{A\,B}-\braket{A}\braket{B}$. 

\bibliography{reference}	
\bibliographystyle{apsrev4-2}

\end{document}